\documentclass[12pt]{article}
\usepackage{geometry} 
\usepackage{multirow}

\usepackage{graphicx}
\usepackage{amssymb}
\usepackage{epstopdf}
\usepackage{authblk}

\title{Analysis of Charm Pair Production at the LHC}

\author[1]{D. Souza}
\author[2]{N. H. Brook}
\affil[1]{\small{Department of Physics, University of Bristol, England.}}
\affil[2]{\small{Department of Physics and Astronomy, University College of London, England.}}
\date{}

\begin{document}

\maketitle
\begin{abstract}
The DGLAP and CCFM approaches to perturbative QCD evolution have been investigated by examining correlations of charmed hadron pairs in $pp$ collisions at $\sqrt{s}=7$~TeV. The theoretical models are compared to the data taken by the LHCb experiment. Differences in the parton kinematics between the two approaches are discussed. In general a model incorporating NLO diagrams matched to parton showers describes the data best.
\end{abstract}

\section{Introduction}

Recent investigations in double $J/ \psi$ and open charm production cross sections in high-energy hadronic collisions have resulted in a growing interest for theoretical and experimental investigations on how pairs of heavy flavoured quarks correlate in their final states \cite{Baranov:2011ch}. This is often viewed as a good environment for understanding multi particle QCD dynamics \cite{Novoselov:2011ff}. 

The LHC is a high luminosity gluon collider and this can be observed in the production of quarkonia. 
The model independent measurements of absolute cross-sections of $J/\psi$ mesons accompanied by open charm hadrons as well as  open charm hadron pairs have already been presented by the LHCb collaboration~\cite{Aaij:2012dz}. The kinematic properties of events containing $J/\psi C$, $CC$ and $C\bar{C}$, has been also compared with theoretical predictions \cite{Berezhnoy:2012xq}. Although there is no discrepancy between the experimental measurements and theoretical calculations, the kinematics of the hadronized charm quarks into $D$ mesons and other physical effect are not completely understood.

In this paper a comparison is presented between different QCD evolution equations containing $C\bar{C}$ pairs and compared with experimental measurements reported by LHCb. The correlations between pairs of charmed mesons are studied and related to the kinematics governing the behaviour of the produced heavy flavour quarks.

\section{Measurements and Model Description}
\label{sect:2}

Measurements of charmed mesons pairs observed at $\sqrt{s}=7$ TeV were reported for first time by the LHCb collaboration. The measurements of open charm cross sections using data collected by the LHCb with an integrated luminosity of $355\pm13 pb^{-1}$ of data during the first half of 2011 gave consistent results with theoretical calculations~\cite{Berezhnoy:2012xq}. 

There are several generators available for heavy quark production that incorporate different methodologies and these have been studied within this paper. \texttt{PYTHIA} is a multi-purpose event generator that uses on mass shell matrix element to generate heavy quark pairs in approach based on collinear factorisation. \texttt{CASCADE} uses off-shell matrix elements and uses unintegrated gluon distributions in the $k_T$ factorisation approach. \texttt{POWHEG} use NLO matrix elements and the cross section is calculated to order $\alpha_s^3.$ The cross section in \texttt{POWHEG} has contributions from the leading order (Born) cross section, one-loop corrections and real-emission contributions as well as the necessary subtractions including those from the initial-state collinear singularities.


Predictions of scaling violations in the parton density functions are known to be dependent on the evolution equations. In this context, the Dokshitzer, Gribov, Lipatov, Alterelli, Parisi (DGLAP) equations which are described in detail in \cite{Altarelli:1977zs} gives us an approximation of the scale evolution for quark and gluon parton densities. In this analysis we used \texttt{PYTHIA 6.4} program \cite{Sjostrand:2006za} to provide a framework to generate high-energy physics events and model the effects the gluon evolution has on the hadronic final state. Further analysis, included generating exact parton level NLO diagrams using the \texttt{POWHEG BOX} tool \cite{Alioli:2010xd} and then hadronization with the same version of \texttt{PYTHIA}, has been undertaken. The formalism that matches NLO QCD to the parton shower is described in \cite{Nason:2004rx}.

Although the DGLAP approximation provides us with a good description of leading logarithms of transferred momenta $\alpha_s \log(Q^2/\mu^2)$, at small-$x$ leading logarithmic terms of longitudinal momenta $(\alpha_s \log x)^n$ are expected to be important \cite{bfkl_1, bfkl_2}. These small-$x$ leading logs should also play an important role in describing the hadronic final state. In this paper the Ciafaloni, Catani, Fiorani, Marchesini (CCFM) approach is used to study these effects \cite{ccfm_1, ccfm_2} as implemented in the \texttt{CASCADE} Monte Carlo event generator \cite{Jung:2010si}. The \texttt{CASCADE} event generator also uses the Lund string model, as implemented in \texttt{PYTHIA}, to model hadronization. 
 
  \begin{figure}[!h]
       \includegraphics[scale=.23]{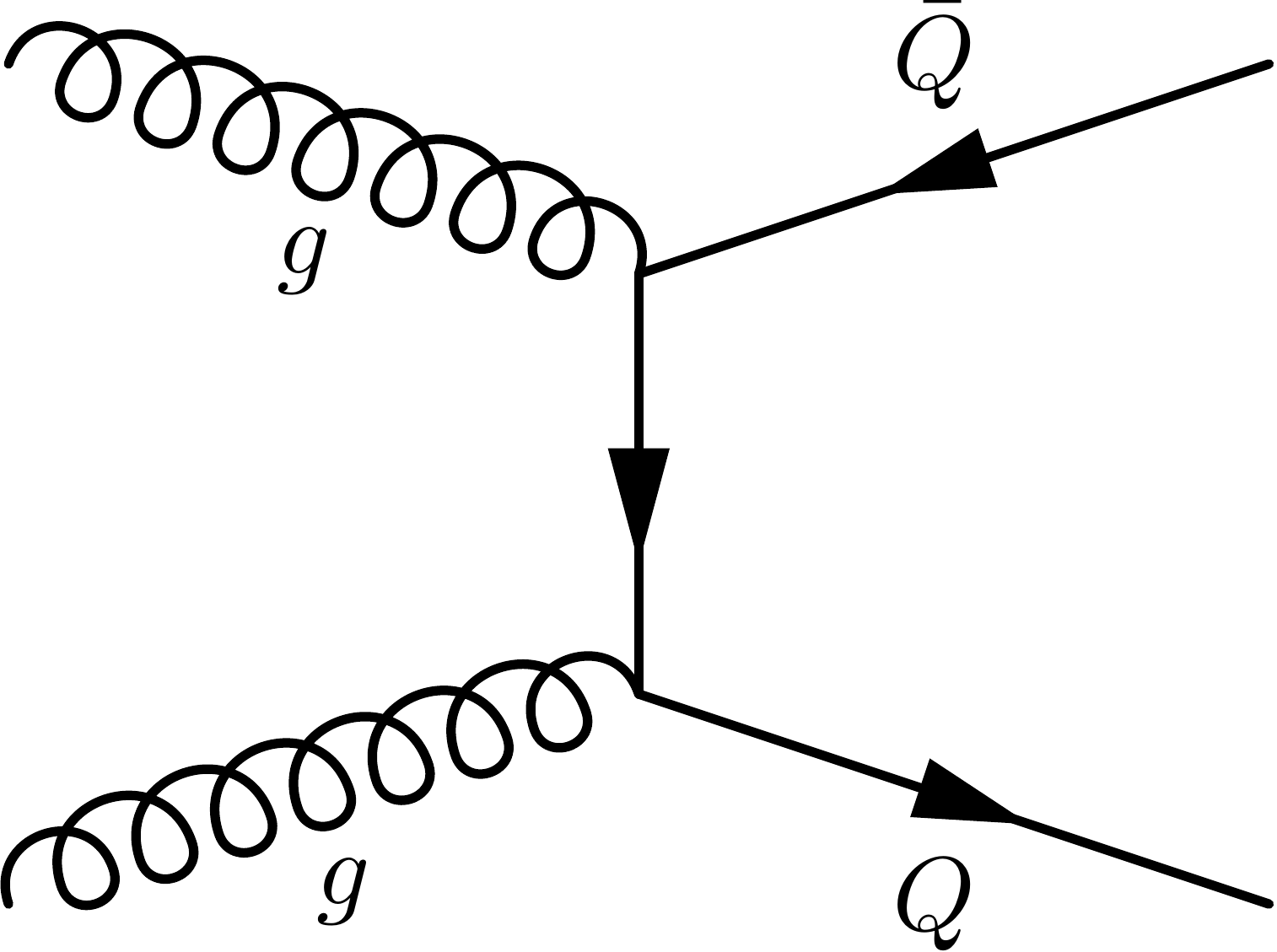} (a)
       \includegraphics[scale=.23]{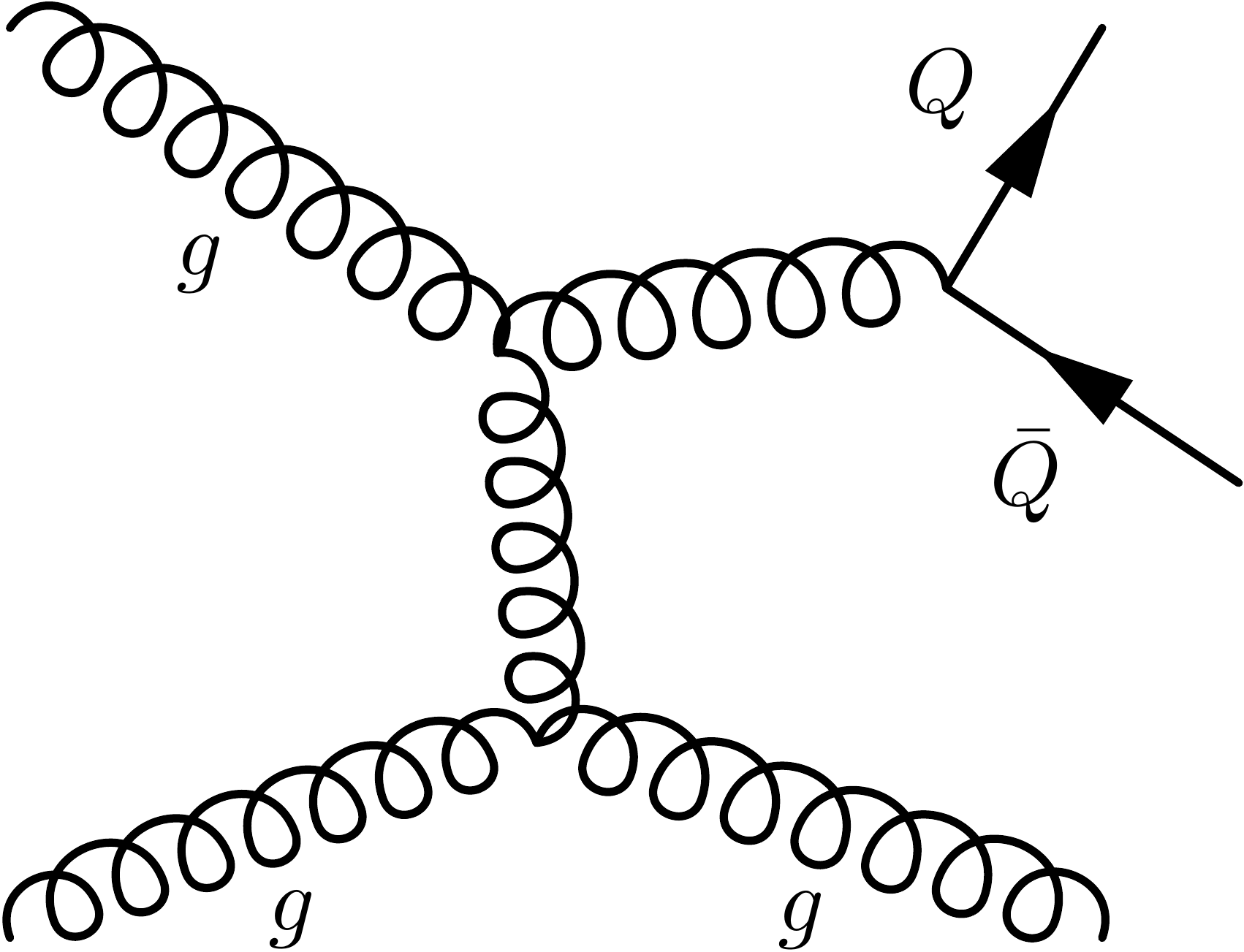} (b)
       \includegraphics[scale=.23]{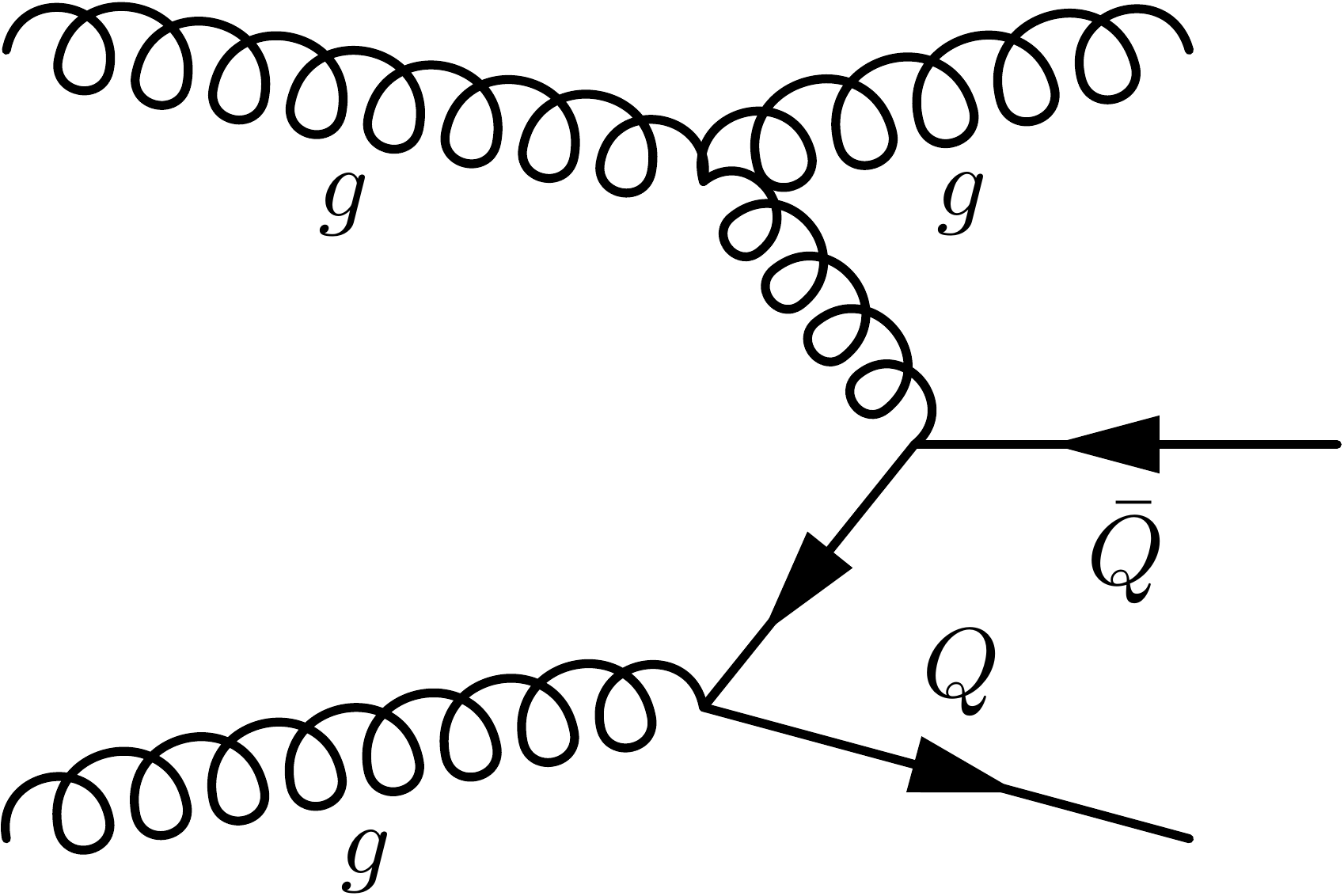} (c)
      
       \caption{Example of three different types of hard scattering, Pair Production (PP), Gluon Splitting (GS) and Flavour Excitation (FE), shown in diagrams (a), (b) and (c) respectively. Note that in a $pp$ collision, the $Q$ quark is not a valence flavour therefore it \emph{must} come from the sea as driven by the gluon content of the proton as illustrated in (c).} 
  \label{fig_dia}
  \end{figure}
  
At Leading Order (LO) of $\mathcal{O(}\alpha_s^2)$, there are several contributions from different heavy flavour production mechanisms to the cross sections, dominated by the Pair Production (PP) processes. The pair produced quarks are generated predominantly by $gg\to Q\bar{Q}$  (Fig. \ref{fig_dia}a). In this process the $Q$ and $\bar{Q}$ have to emerge back-to-back in azimuth angle ($\phi=180^0$) to conserve transverse momentum. There are other next-leading-order (NLO) perturbative processes that can contribute to heavy quark production, including the graphs $gg\to Q\bar{Q}g$ which can disturb the 2 quarks from being back-to-back. There are additional subprocesses of $\mathcal{O(}\alpha_s^3)$ that can be classified.  There is Gluon Splitting (GS) and Flavour Excitation (FE). A GS event is when a gluon is produced that then splits to $g\to Q \bar{Q},$ where no heavy flavours enter the original LO hard scattering (Fig. \ref{fig_dia}b). The FE process occurs when a heavy flavour from the parton distribution of one of the incoming proton beams is put on mass shell by scattering a parton from the other beam (Fig. \ref{fig_dia}c). It should be noted that \texttt{PYTHIA} can generate GS, PP or FE processes while \texttt{CASCADE} considers only PP and GS. 
 
 \texttt{CASCADE}, uses the CCFM approach to parton showers, which bridges between the DGLAP and BFKL approaches at small-$x$. In this approach the transverse momentum component of the incoming partons is treated explicitly. This means that one can have a transverse momentum of the gluon which is larger than the one of the outgoing quarks. Such situation can also be obtained from a NLO diagrams of charm production such as $gg\to c\bar{c}+g$. In contrast, in the DGLAP approach, gluons are the result of parton evolution from the proton, where the partons (from the shower) are generated in a collinear approximation.
 
\section{Results}

The kinematic distributions for double charm production were generated for different pairs of D-mesons and correlated as a function of the transverse momentum $P_T$, difference in rapidity $|\Delta y|$ and in azimuthal angle $|\Delta \phi|$ between the two hadrons using all 3 event generators. The difference in rapidity or azimuthal angle are defined as
\begin{equation}
   | \Delta y | = |y_{h} - y_{\bar{h'}}| \\
\end{equation}
\noindent or
\begin{equation}
  |  \Delta \phi | = |\phi_{h} - \phi_{\bar{h'}}|
\end{equation}
respectively, where $h$ stands for the following hadrons : $D^0$, $D^+$, $D_s^+$, $\Lambda_c^+$ (or corresponding anti-particle $\bar{h'}$). 

The event generator studies explore the same kinematic region as the LHCb analysis,  $3<P_T<12$GeV/c and $2.0<\eta<5.0$ where $\eta$ is the particle's pseudorapidity.

\begin{table}[h!]
\centering
\begin{tabular}{l|l|l}
	\hline
	\hline
	Case & Subprocess & Interactions \\ \hline
	\rule{0pt}{12pt} 
	\multirow{2}{*} 1 & \multirow{2}{*}{GS only} & $gg \to gg$ \\ 
	 && $f\bar{f} \to gg$ \\
	\hline
	\rule{0pt}{12pt} 
	\multirow{2}{*} 2 & \multirow{2}{*}{PP or GS} &  $gg \to f\bar{f}$ \\ 
	 && $f\bar{f} \to f'\bar{f}'$ \\
	\hline
	\rule{0pt}{12pt} 
	\multirow{2}{*} 3 & \multirow{2}{*}{FE or GS} & $fg \to fg$ \\ 
	 && $ff' \to ff'$ \\
	\hline
	\hline
	\end{tabular}
	\bigskip
\caption{Different flavour ($f$), anti-flavour ($\bar{f}$) and gluon ($g$) interactions considered in the analysis grouped in three pairs. In the first case, $c\bar{c}$ pairs are generated exclusively via GS while in the second and third cases they can also be produced via PP and FE respectively. Note that in all 3 cases GS could also happen by gluon radiation generated in subsequent parton showers.}
\label{table1}
\end{table}

Table \ref{table1} shows a summary of the processes involved in heavy quark production and their classification as outlined in section~\ref{sect:2}. In the first case, the required $c \bar{c}$ pair is always produced via gluon splitting by either of the outgoing gluons. Note that any final state partons with enough energy to produce gluons can also result in a $c\bar{c}$ pair being produced. 

The second case, considers two interactions where LO fusion processes are  dominant, either by incoming gluons but also by an incoming quark and anti-quark. In this case the outgoing partons can be a $c\bar{c}$ pair. Also GS processes can contribute since the final state parton can radiate a gluon with enough energy to produce a $c\bar{c}$ pair even if non-charm flavours were produced in the LO process. The third case is the flavour excitation process. Here an incoming quark, which comes from one of the incoming protons, interacts with a gluon from the other proton beam and is put on mass shell. This quark can either be $c$ or $\bar{c}$. In the case on incoming $ff'$, only one quark is required to be a charm quark. Double flavour excitation is also possible although very improbable. As in the other cases, GS processes can also contribute since a final state parton can radiate a gluon with enough energy to produce a $c\bar{c}$ pair.

  \begin{figure}[!h]
  \centering
       \includegraphics[scale=.46]{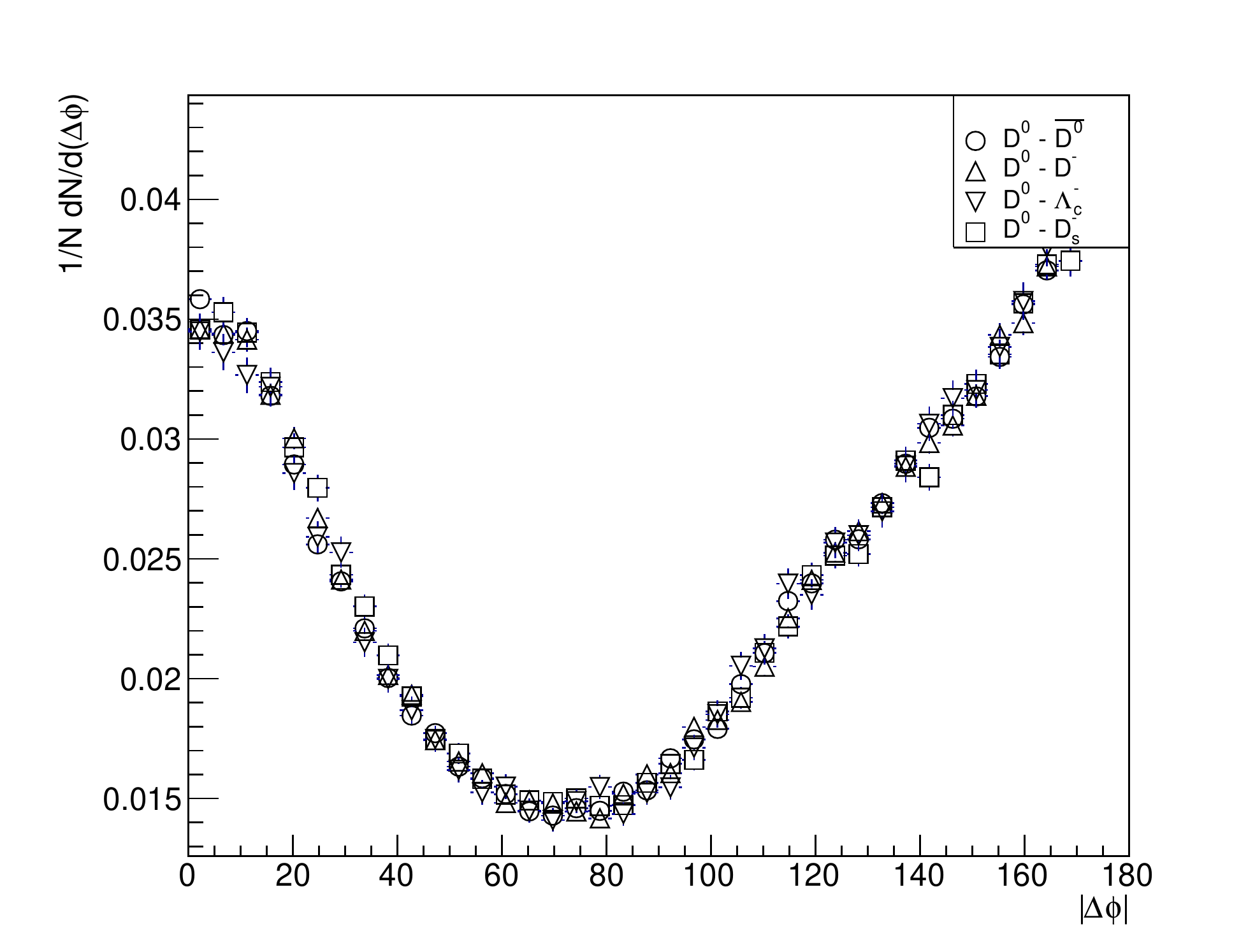}
       \caption{Normalised \texttt{PYTHIA} MC prediction of the azimuthal angle spectrum for all the studied charmed hadrons pairs. }
       \label{dphi_py}
  \end{figure}

In the azimuthal angle spectra, Figure \ref{dphi_py}, the \texttt{PYTHIA} generator has two peaks located near $|\Delta \phi| = 0, 180$. Different pairs of charmed mesons show to have similar behaviour therefore only $D^0, \bar{D^0}$ pairs are shown in subsequent figures.

\begin{figure}[!h]
    \centering
       \includegraphics[scale=.46]{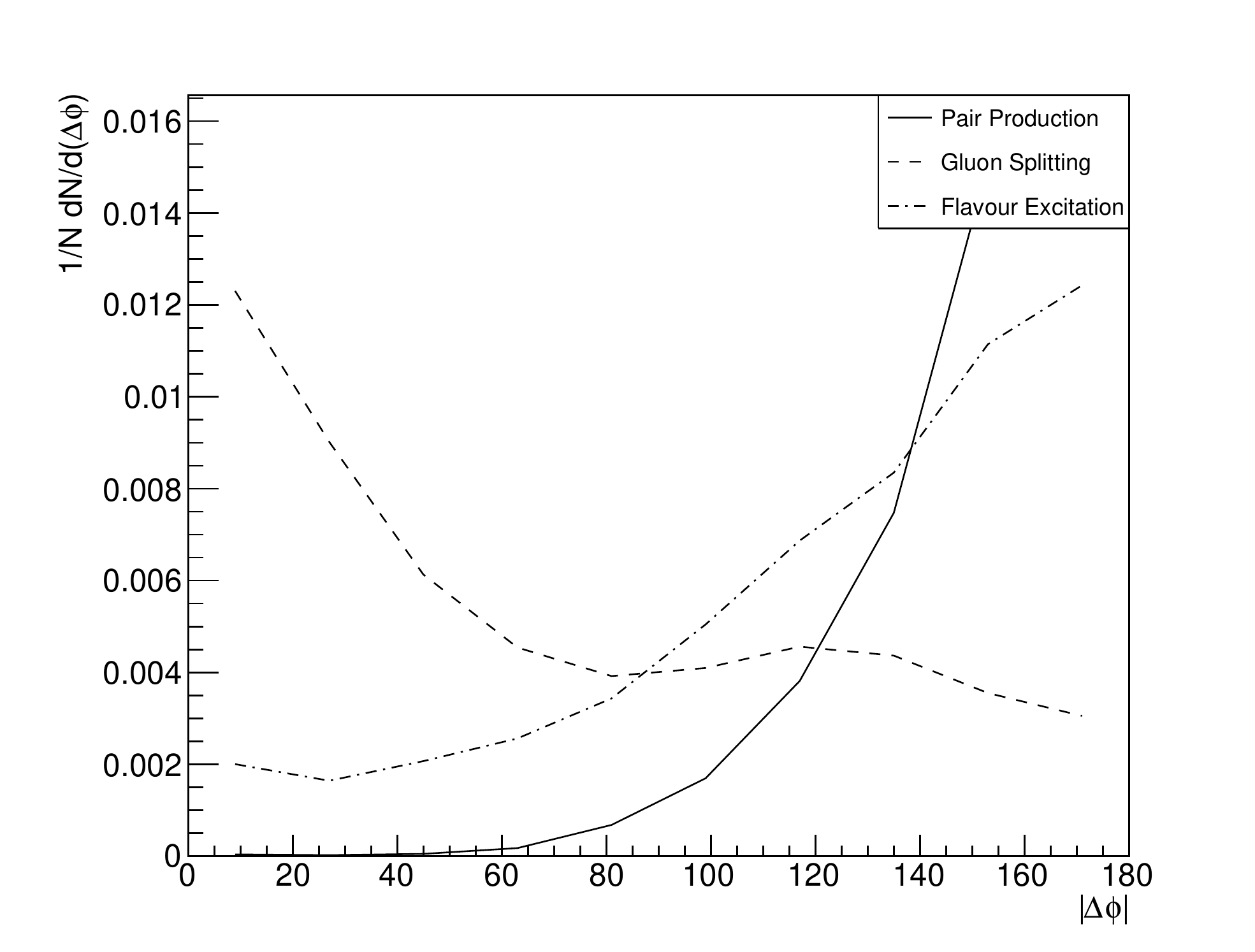}
       \caption{Normalised \texttt{PYTHIA} $\Delta \phi$ distribution of the three parton level processes PP (solid line), GS (dashed line) and FE (dot-dashed lined).} 
       \label{pp_gs_fe}
  \end{figure}

To understand better the shape of the distribution, the individual parton level contributions from each of the three subprocesses outlined in Table 1, were investigated. As observed in Figure \ref{pp_gs_fe} the gluon splitting contribution peaks at low values of $|\Delta \phi|$ as expected for pairs created from a single gluon, whilst quark pairs  via PP show a clear peak near $180^0$ and pairs produced through flavour excitation also tend to be more back-to-back.

\begin{figure}[!h]
  \centering
       \includegraphics[scale=.46]{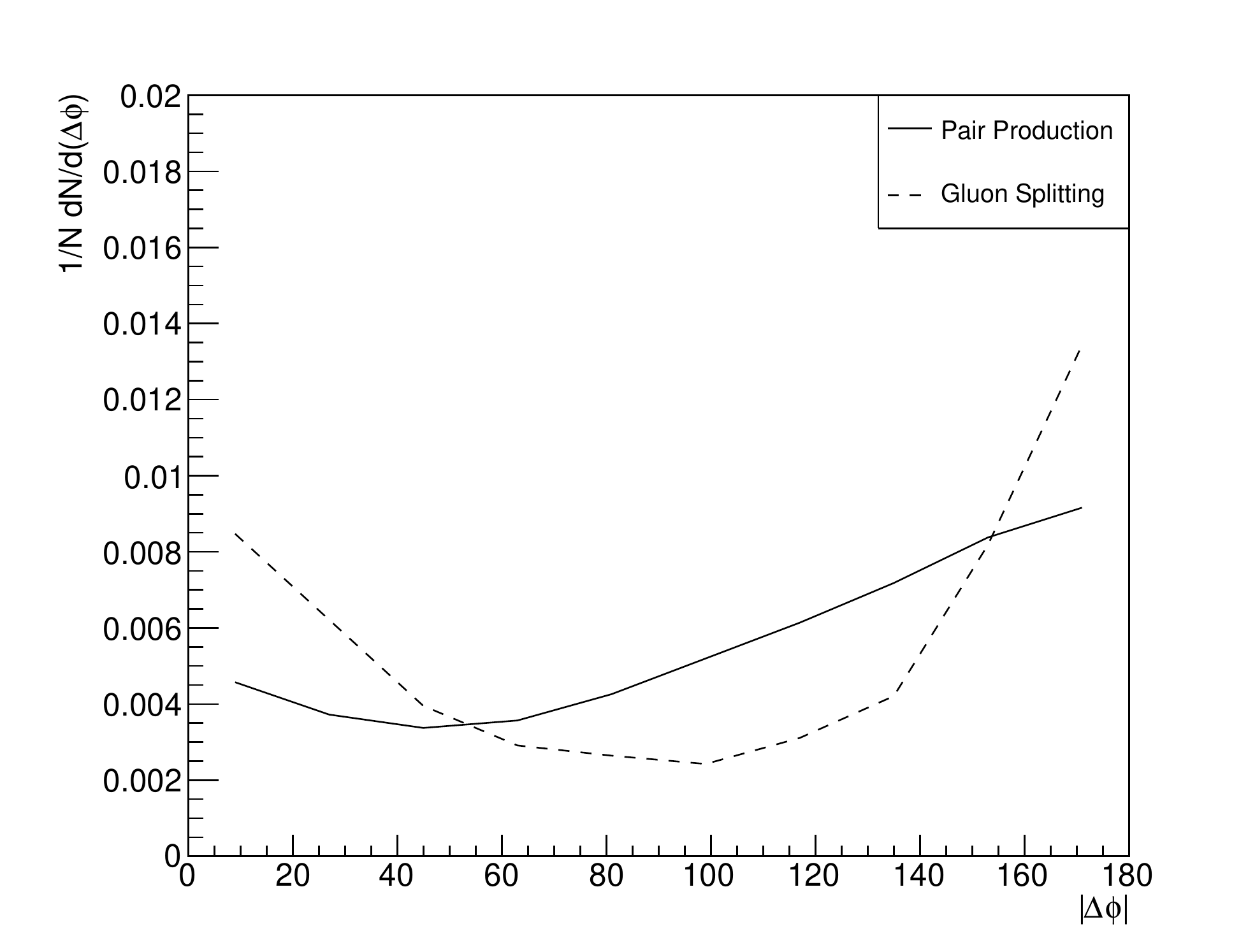}
       \caption{Normalised \texttt{CASCADE} spectra of $\Delta \phi$ showing PP (solid) and GS (dashed) contributions. } 
       \label{pp_gs_k}
  \end{figure}

The same parton level study was performed for the \texttt{CASCADE} generator, Figure \ref{pp_gs_k}, but surprisingly this time the kinematic distribution for the  PP and GS processes differ from that observed for \texttt{PYTHIA}. For \texttt{CASCADE} the PP hard scattering kinematics dominates giving a peak, as expected from the LO, not only when $|\Delta \phi| \to 180$ but also near zero. 
This is due to the lack of $P_T$ ordering in CCFM approach to the evolution of the parton shower that results in  the gluons from the incoming proton beams carrying high values of transverse momentum components perpendicular to the beam axis, creating a \emph{momentum kick}. 

\begin{figure}[!h]
    \centering
       \includegraphics[scale=.46]{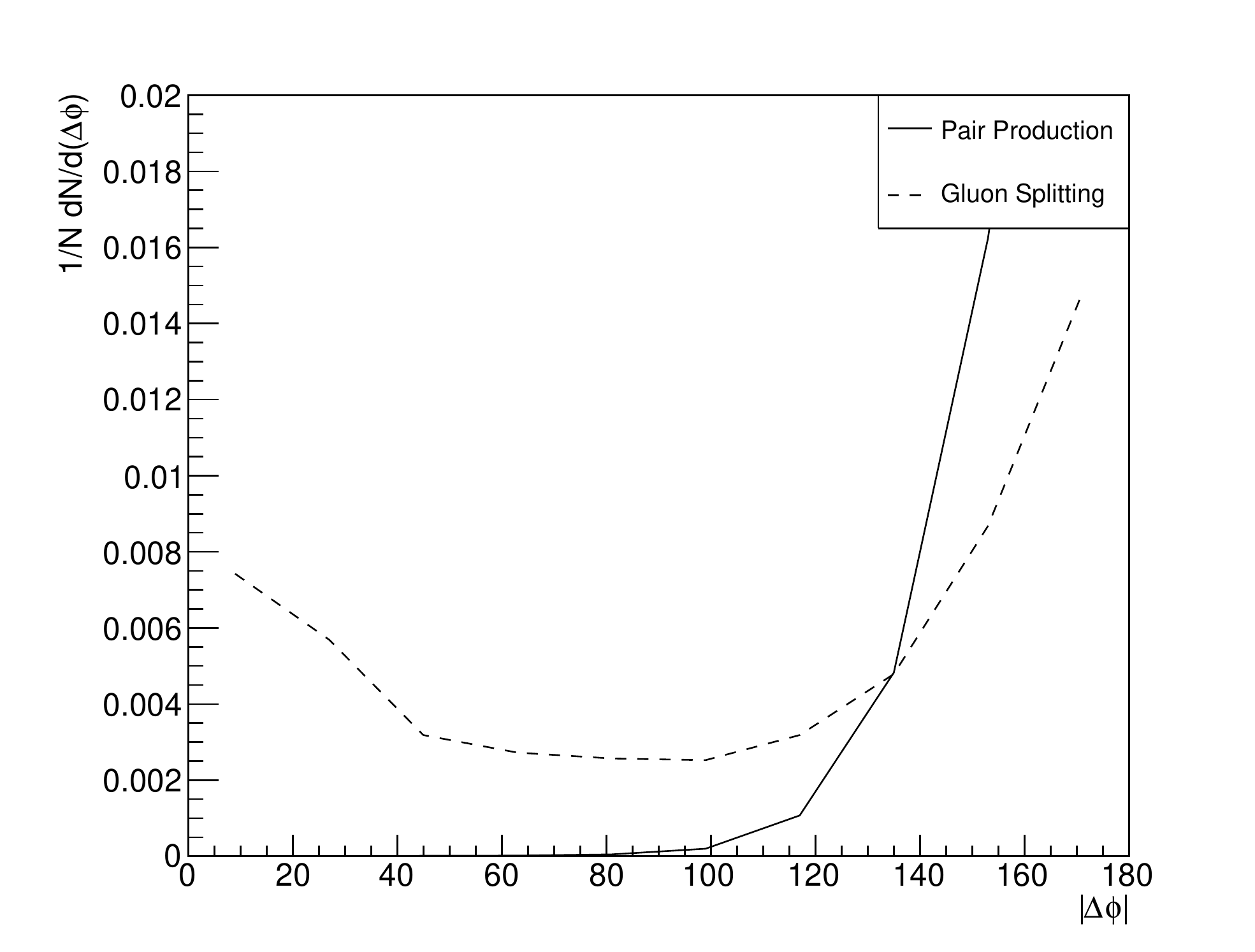}
       \caption{Normalised \texttt{CASCADE} spectra of $\Delta \phi$ after applying momentum cut at parton level, showing PP (solid) and GS (dashed) contributions. } 
       \label{pp_gs_nk}
  \end{figure}

By applying an (arbitrary) transverse momentum cut at the parton level of $2.5$GeV for incoming gluons in the CCFM approach, those events where the gluon is larger than the momentum of the outgoing quarks are discarded. Figure \ref{pp_gs_nk} illustrates the effect of such a cut. This is a consequence of the leading logarithmic terms of longitudinal momenta implemented in CCFM and is not seen within the DGLAP evolution equations (\texttt{PYTHIA}) where the GS at high $|\Delta \phi|$ shows a flat behaviour, see solid line in Figure \ref{pp_gs_fe}.

With the transverse momentum cut on \texttt{CASCADE} the peak at $\Delta \phi  \approx 0$ now disappears and the results exhibit similar behaviour for PP events in \texttt{PYTHIA}. For \texttt{CASCADE} the GS case, a rise is seen in the small and large $|\Delta \phi|$ regions. The peak for low values of $|\Delta \phi|$, is also seen in Figure \ref{pp_gs_fe} for GS processes in \texttt{PYTHIA}, and its kinematics can be explained as energetic gluons splitting into charm-anticharm pairs that are (transversely) boosted to the lab frame with small opening angles.

An additional peak is observed for large angular differences in both cases, with and without \emph{momentum kicks}. In those events, the $c\bar{c}$ pair is produced by gluons with a large longitudinal momentum component almost parallel to the beam line, where the polar angle $\theta$ is nearly zero. This results in a (transverse) back-to-back decay as if they were produced in the rest frame. 

\begin{figure}[!h]
         \centering
       \includegraphics[scale=.46]{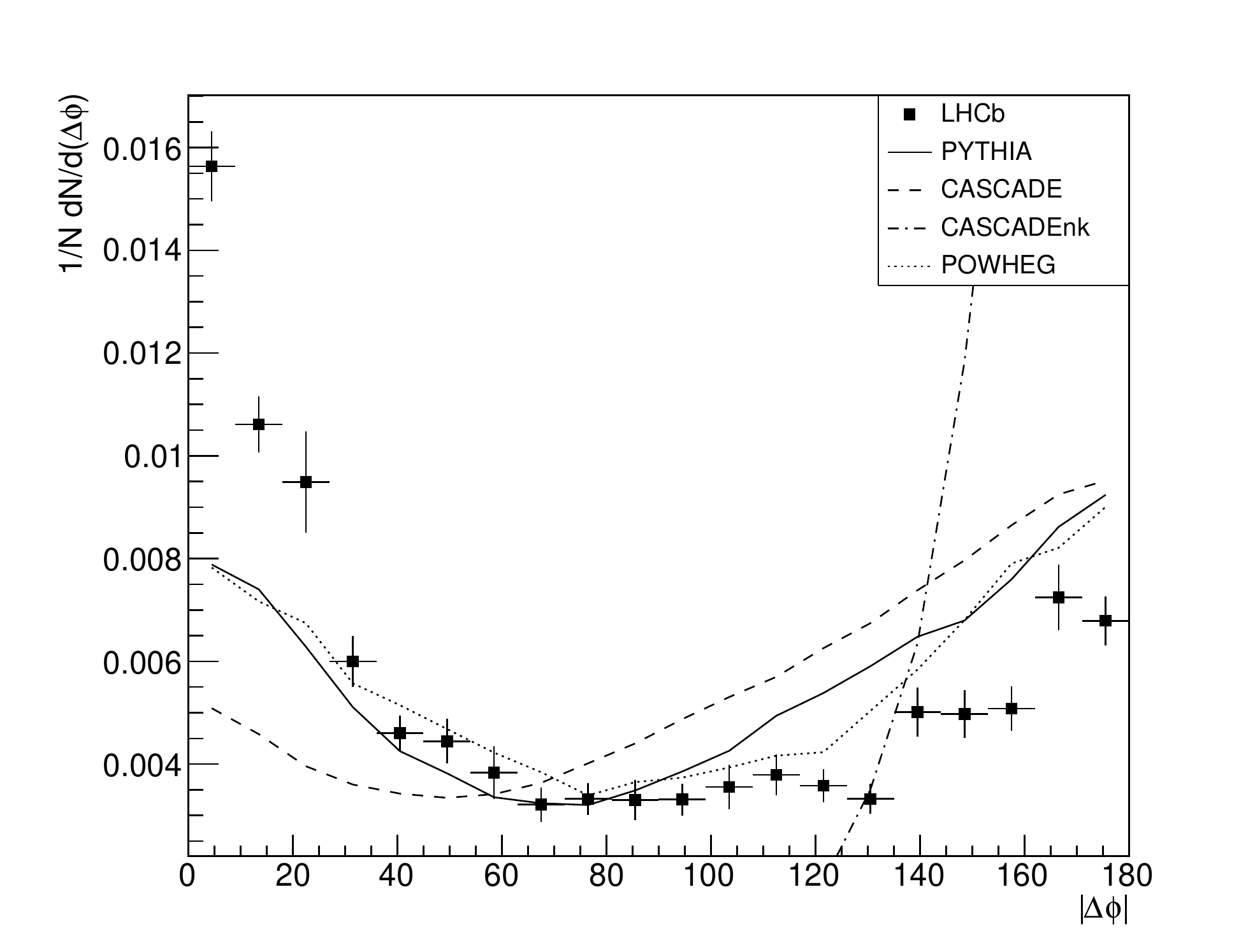}
       \caption{Comparison of event generator (\texttt{PYTHIA}, \texttt{CASCADE} and \texttt{POWHEG}) azimuthal angle difference ($|\Delta \phi|$) between $D^0$ and $\bar{D^0}$ mesons and LHCb data.}
        \label{sick}
  \end{figure}

A normalised plot showing the azimuthal angular correlationn between $D^0,\bar{D}^0$ pairs is shown in Figure \ref{sick}, where the solid squares correspond to LHCb data and the lines to the different generator models. 
There is agreement between data, LO and NLO DGLAP description (\texttt{PYTHIA} and \texttt{POWHEG} respectively) for high values of $|\Delta \phi|$ where pair production processes are expected to dominate. In this same region the CCFM description excluding \emph{momentum kicks} peaks much higher than data. For the \texttt{CASCADE} model the \emph{kick} and \emph{no-kick} (nk) cases are illustrated separately.

None of the theoretical models agree completely in the description of either the high or the low $|\Delta \phi|$ values. One possible reason that the experimental data peaks higher at low $|\Delta \phi|$ may also be due to the presence of the charmonium state $\psi(3770)$, which is near threshold for $D^0, \bar{D}^0$ decay and in the laboratory frame will be boosted to $\Delta \phi \approx 0.$ 
A comparison of the quality of all models is summarised in Table \ref{table2}.

\begin{table}[h!]  
\centering
\begin{tabular}{l||l|l|l|ll|l}
	\hline
	\hline
	Variable & \texttt{PYTHIA} & \texttt{CASCADE} & \texttt{CASCADE}(nk) & \texttt{POWHEG} \\ \hline
	\hline 
	$\Delta \phi$ & 20.3 & 44.7 & 268.1 & 14.2 \\ \hline
	$P_T$ & 12.50 & 1.52 & 37.0 & 1.42 \\ \hline
	$|\Delta y|$ & 3.4 & 12.4 & 25.2 & 17.2 \\ \hline
	\hline
	\end{tabular}
	\bigskip
	\caption{Summary of the $\chi^2/dof$ between LHCb data and each studied model.}
	\label{table2}	
\end{table}
  
\begin{figure}[!h]
    \centering
       \includegraphics[scale=.46]{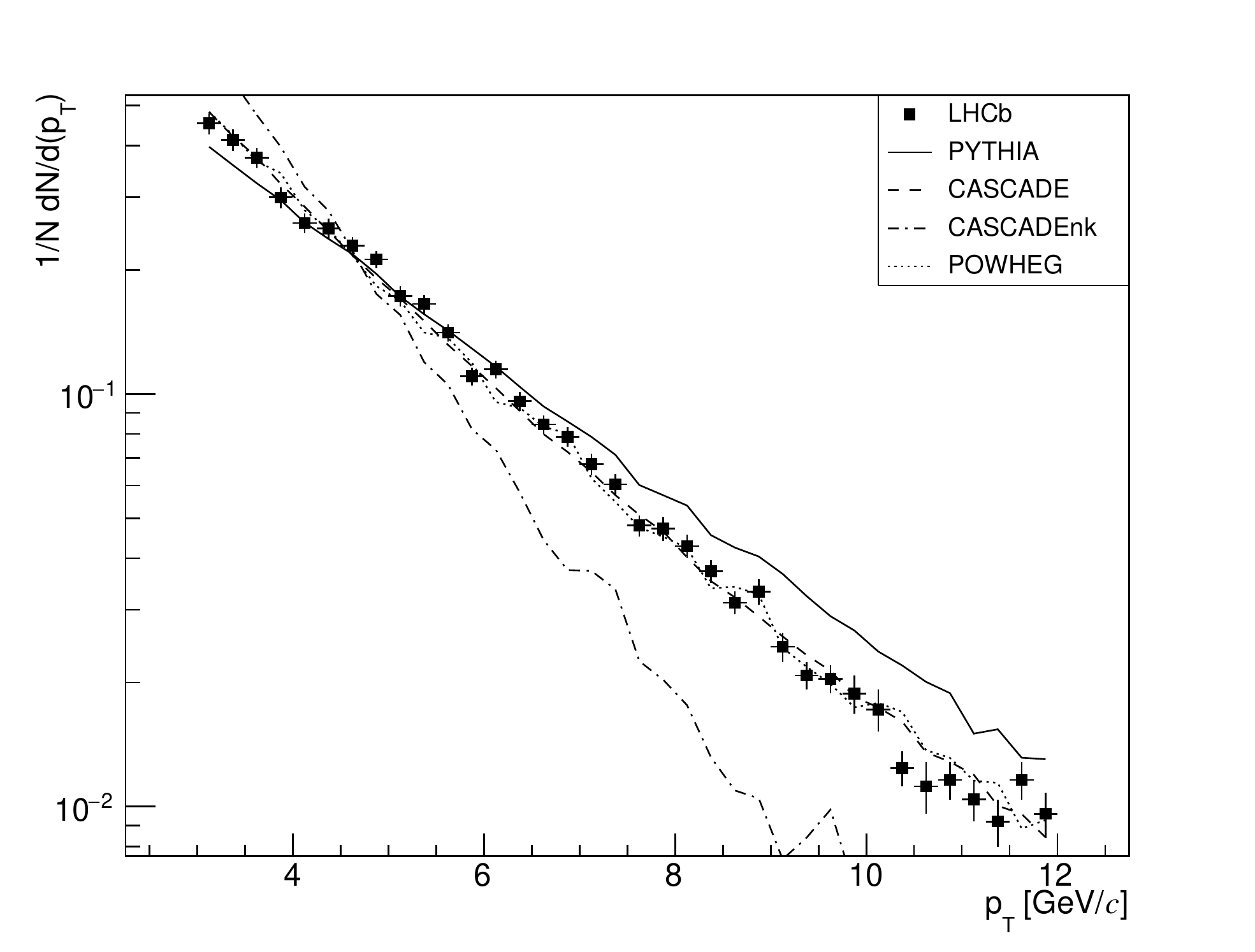}
       \caption{LHCb data compared to \texttt{PYTHIA}, \texttt{CASCADE} and \texttt{POWHEG} models. The $P_T$ spectra predictions showed correspond to $D^0$ and $\bar{D}^0$ mesons.}
       \label{pt_plot}
  \end{figure}

The transverse momentum spectra is shown in Figure \ref{pt_plot}. The four generator models  are compared using a sample of $D^0$ particles. \texttt{PYTHIA} (solid line), \texttt{CASCADE} with no gluon \emph{momentum kick} cut (dashes) are in good agreement. Also the \texttt{POWHEG} result (dotted line) shows a good agreement with the data. Not unexpectedly if the \emph{momentum kicked} events from \texttt{CASCADE} are removed (dash-dotted line), the $P_T$ spectra slope drops dramatically. The quantitative level agreement is given in Table \ref{table2}.

\begin{figure}[!h]
    \centering
       \includegraphics[scale=.46]{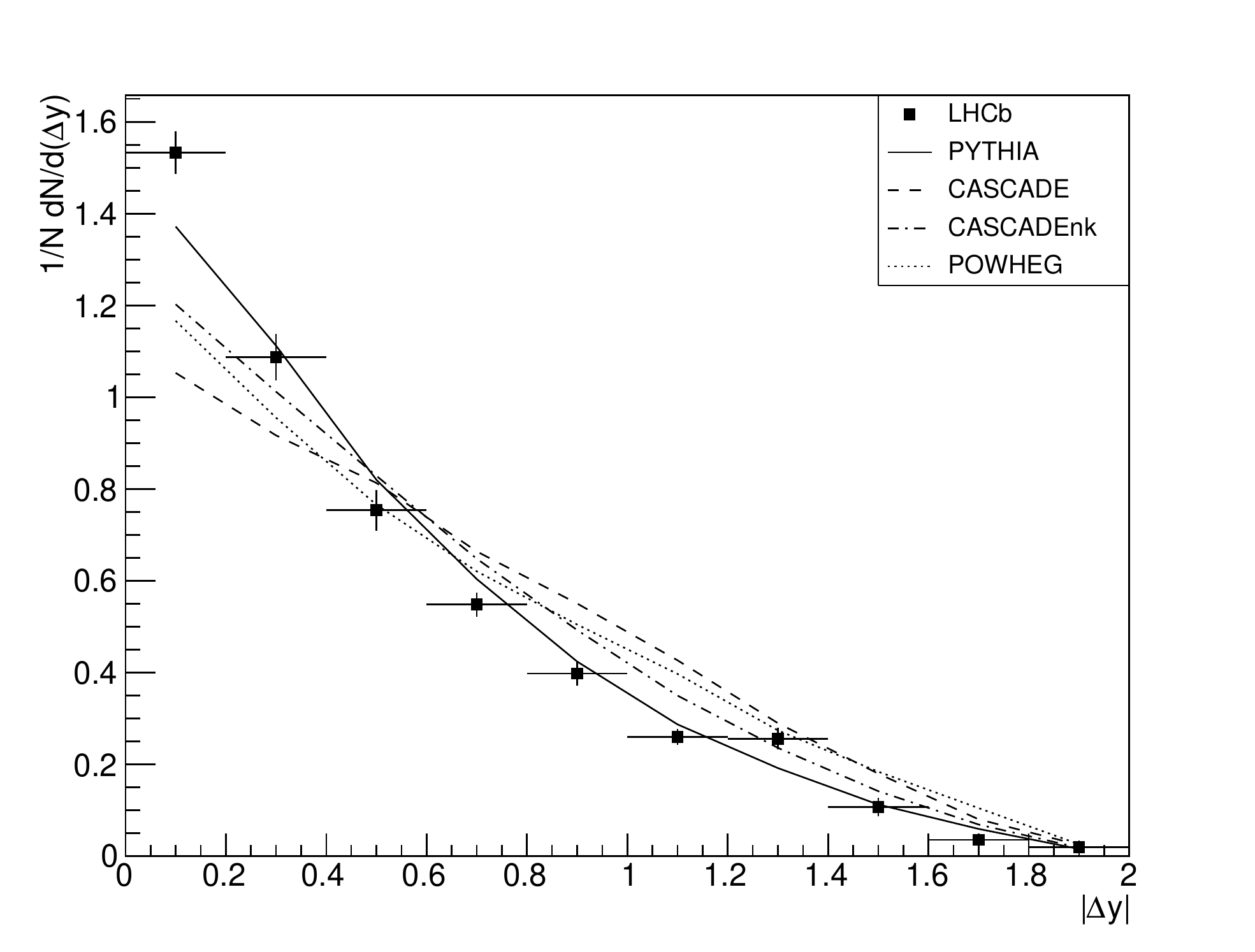}
       \caption{Comparison of event generator (\texttt{PYTHIA}, \texttt{CASCADE} and \texttt{POWHEG}) rapidity difference between $D^0$ and $\bar{D^0}$ mesons and LHCb data.}
           \label{dy_plot}
  \end{figure}

Figure \ref{dy_plot} shows the differences between the distributions in $|\Delta y|$ for the event generators with a $D^0\bar{D}^0$ pair sample. Broader $|\Delta y|$ is observed in the data than in the event generators. This is suggestive of contributions to double parton scattering being present in the data \cite{Kom:2011bd}. None of the generators studied simulated Multi-Parton Interactions (MPI). 
Although \texttt{PYTHIA} has a MPI option, this does not consider 2 pairs of charm partons being produced, excluding the possibility of several $c\bar{c}$ pairs being generated so unsuitable for further study in this context.

After comparing the three generators, one can conclude that the best description is done by \texttt{POWHEG} followed by \texttt{PYTHIA} and then \texttt{CASCADE}. This is summarised in Table \ref{table2}. 
It is important to remark that the calculation of the $\chi^2$ between the models and LHCb data was done considering both statistic and systematic errors.

\section{Summary}

Measurements of $P_T$, $|\Delta y|$ and $|\Delta \phi|$ spectra for several charmed hadron pairs in particular  $D^0,\bar{D}^0$ observed at LHCb were compared with different theoretical models of QCD evolution, DGLAP and CCFM.

The results show that gluon splitting processes contribute to low angle increases in the $\Delta \phi$ spectra. Although this is theoretically expected it is not the only process that makes a contribution in the $\Delta \phi \to 0$ range. 

These studies suggest that pair produced charms are also sensitive to the underlying QCD evolution of the partons within the proton.

\section*{Acknowledgements}

Authors would like to thank Hannes Jung for his fruitful explanations of CCFM dynamics and A. Novoselov \& team for their fast reply, support and interesting interpretation of $P_T$ spectra in gluon collinear approaches. This work was supported by CONICYT Chile under the grant number 72120187.

\bibliographystyle{utphys}  
\bibliography{Bibliography}  

\end{document}